# PEDRO-V: FROM A CONCURRENT ENGINEERING CASE STUDY TO A PROMISING PHASE ZERO MISSION DEFINITION

**Domenico D'Auria** [1,2], **Arianna Rigo**[3], **Luca Niero**[4*], **Andrei-Toma Stoica**[4*], **Vito Costantini**[4*], **Pasquale Castellano**[4*], **Zsofia Zita Szilagyi**[4*], **Nishani Vijayakumaran**[4*], **Ella Toppari**[4*], **Stefano Schiano**[4*], **Marco Adorno**[4*], **Matteo Matrone**[4*], **Chiara Tulli**[4*], **Jan Kurowski**[4*], **Leo Bougault**[4*], **Argenziano Francesco**[4*], **Antignano Claudia**[4*], **Theodoros Roumanis**[4*], **Victoria Koßack**[4*], **Spyridon Giuvalas**[5]

[1]*Istituto Nazionale di Astrofisica – Osservatorio Astronomico di Capodimonte* / [2]*Università degli Studi di Napoli "Federico II" – Dipartimento di Ingegneria Industriale*
*Salita Moiariello, 16, 80131, Naples / Via Claudio, 21, 80125 Naples, Italy*
*Email: domenico.dauria@inaf.it*

[3] *Instituto Superior Técnico – Universidade de Lisboa*
*Instituto Superior Técnico, Av. Rovisco Pais 1, 1049-001 Lisbon, Portugal*
*Email: arianna.rigo@tecnico.ulisboa.pt*

[4*] *Other institutes/universities (see acknowledgments)*

[5] *University of Luxembourg*
*2 Avenue de l'Universite L, 4365 Esch-sur-Alzette, Luxembourg*
*Email: spyridon.gouvalas@uni.lu*

**INTRODUCTION**

The rapid advancement of space exploration necessitates innovative approaches to mission design that can significantly reduce development time and cost. Concurrent engineering (CE) has emerged as a crucial methodology in the space industry, addressing these needs by facilitating collaborative and interdisciplinary design processes. The CE approach ensures that complex space missions are carried out efficiently, leveraging the expertise of various engineers to produce optimal design solutions within a condensed time frame [1, 2]. This paper presents the final design of the PEDRO-V mission's Phase 0/A feasibility study, which explores Venus' atmosphere, as part of the ESA Academy's Concurrent Engineering Challenge 2024. The challenge, engaging students from multiple European universities, simulated a professional collaborative engineering environment, emphasizing the educational and practical benefits of concurrent engineering. The detailed PEDRO-V mission design illustrates CE methodologies in an academic setting, showcasing iterative design sessions, subsystem developments, and trade-off analyses. The involvement of ESA experts ensured mission objectives were met, demonstrating CE's ability to foster innovative concepts and streamline space mission design among student teams. This paper focuses on the final design of a Phase 0/A feasibility study of the PEDRO-V mission, aimed at exploring Venus' atmosphere, and its designing process. The study was part of the ESA Academy's Concurrent Engineering Challenge 2024, a program designed to engage students in the complexities of concurrent design. The primary contribution of this paper is the detailed presentation of the PEDRO-V mission design, which showcases the efficacy of CE methodologies in an academic environment. The paper outlines the iterative design sessions, subsystem developments and trade-off analyses conducted by the students. This research underscored the educational benefits and practical applications of CE in developing feasible space missions, demonstrating how it can foster innovative concepts among non-expert student teams and streamline the mission design process. The paper explores CE in the space sector, discusses integrating Project-Based Learning within CE sessions, details methodologies and tools from the ESA Concurrent Engineering Challenge 2024, highlights the PEDRO-V mission design outcomes, and examines the educational impact, lessons learned, and future research directions.

**Concurrent Engineering**

Concurrent Engineering (CE) is a comprehensive and systematic approach to the integrated and parallel design of products and processes. It involves the simultaneous development of various project activities, fostering continuous information exchange among interdisciplinary teams. The primary goal of CE is to accelerate project completion by allowing multiple aspects of the design and development process to occur concurrently, rather than sequentially [1, 3].

**State of the art**

Historically, space mission design has relied on the traditional sequential engineering approach, following a linear V-model process from defining system requirements through conceptual and detailed design to implementation, verification, and maintenance. Each phase must be completed before the next can begin, ensuring a structured development path but leading to inefficiencies when modifications in later stages necessitate revisiting earlier stages, causing delays and increased costs [3]. In this model, early-phase 0/A feasibility studies under the European Cooperation for Space Standardization (ECSS) standards involve identifying feasible solutions and defining requirements, followed by conceptual design that explores various options through trade-off analyses. While this approach thoroughly examines potential designs, it may not be optimal for complex missions where subsystem changes can significantly impact others, requiring iterative reworks and extended timelines [4]. CE optimizes Phase 0/A feasibility studies by fostering parallel development and continuous interaction among multidisciplinary teams. It systematically integrates project activities concurrently, accelerating project completion by emphasizing early consideration of all life-cycle aspects. Engineers from different disciplines collaborate simultaneously on various project elements, minimizing design iterations and enhancing efficiency and product quality. This approach also reduces development time and costs compared to traditional sequential methods [1, 3]. The European Space Agency (ESA) has been at the forefront of CE adoption, with its Concurrent Design Facility (CDF) at ESTEC playing a pivotal role. Established in 1998, the CDF was one of the first facilities to institutionalize CE for space mission design. It has since facilitated the development of numerous complex missions, providing a platform where interdisciplinary teams can collaborate efficiently using advanced tools and methodologies [5, 3]. CE achieves these benefits through real-time communication and data exchange among team members, allowing for immediate feedback and rapid iteration on design concepts and immediate assessment of changes across the system. Additionally, CE fosters a collaborative environment where diverse expertise converges to optimize system design, promoting innovative solutions and streamlined processes [6, 7].

**Project Based Learning in CE**

Project Based Learning (PBL) is an essential component of engineering education that aligns well with CE principles. PBL involves students in real-world projects that require collaborative problem-solving and interdisciplinary teamwork. In the context of CE, PBL allows students to experience the complexities of space mission design, enhancing their technical skills and fostering a deep understanding of the interconnected nature of engineering disciplines. This educational approach prepares students for professional roles in aerospace engineering by providing practical experience in a simulated industry environment [8, 9].

**The Concurrent Design Facility**

The Concurrent Design Facility supports CE methodologies as a collaborative workspace for engineers from diverse specialties to conduct rapid and integrated conceptual design of space missions. ESA's CDF, located at ESTEC, has a layout and infrastructure that efficiently facilitates communication among subsystems, ensuring rapid and integrated execution of design processes [10].

**COMET, the Concurrent Model-based Engineering Tool**

Since January 2022, the ESA CDF has adopted the Concurrent Model-based Engineering Tool (COMET) [11] to enhance its CE processes. Developed by RHEA Group, COMET is an open-source tool that enables effective data sharing and features a modern user interface. COMET supports near real-time collaboration, allowing team members to synchronize their work and ensure seamless integration of design parameters. Its adoption in ESA Academy's workshops and challenges underscores its importance in both educational and professional environments [5].

**Current Status and Future Directions**

The CDF at ESTEC has been operational for over two decades, continuously evolving to support the latest advancements in CE methodologies. The facility's success in conducting Phase 0/A feasibility studies has paved the way for numerous space missions. A recent survey across industry, government agencies, and academic organizations with CDFs highlighted the benefits of CE, including reduced study duration and improved design quality. However, it also identified challenges such as managing large data volumes and ensuring effective team communication. Future research aims to address these challenges and further refine CE processes to enhance their applicability in the space sector [2].

**CONCURRENT ENGINEERING CHALLENGE 2024**

The ESA Academy's Concurrent Engineering Challenge 2024 was an immersive, hands-on event designed to engage students in the CE process. This challenge brought together 102 students from various ESA member states, divided into four teams, each located at different Concurrent Design Facilities (CDFs) across Europe: the European Space Security and Education Centre (ESEC) in Belgium, the National Higher French Institute of Aeronautics and Space (ISAE-SUPAERO) in France, the National and Kapodistrian University of Athens (NKUA) in Greece, and the University of Portsmouth in the UK. The primary goal of the challenge was to provide students with practical experience in the CE methodology, mirroring the real-world processes used by ESA and other leading space organizations. Over five days, the participants engaged in a rigorous schedule that included iterative design sessions, subsystem development, and trade-off analyses. Each of the four teams was composed of students with diverse backgrounds and expertise, assigned to various satellite subsystems such as Trajectory Analysis, Attitude and Orbit Control System (AOCS), Propulsion, Thermal Control System (TCS), Communications and Data Handling (C&DH), Electrical Power System (EPS), Mechanisms, Structure and Configuration. This multidisciplinary approach ensured that all aspects of the mission design were considered and integrated. Throughout the challenge, the team working at the ESEC facility was supported by ESA system engineers who provided guidance and acted as project managers. Each participant had access to a dedicated workstation equipped with essential software tools, including COMET [11] and specialized spreadsheets developed by ESA Academy and the Polytechnic University of Turin [12]. The challenge was structured to emphasize collaboration and continuous communication. Daily progress meetings were held to review mission design status, address issues, and share insights across the different CDFs. These daily briefings were crucial for maintaining alignment and ensuring steady progress of all teams towards shared objectives. The final day of the challenge featured a comprehensive review session, where each team presented their mission designs and discussed the engineering decisions made throughout the week. This collaborative review not only allowed for the comparison of different approaches but also provided valuable learning opportunities, highlighting the strengths and potential improvements in each design.

**MISSION DEVELOPMENT AND SUBSYSTEM DESIGN**

The students were tasked with defining a comprehensive mission focused on planetary exploration, culminating in the proposal of the PEDRO-V mission (Planetary Exploration Deployment and Research Operation – Venus). Inspired by the Soviet Union's Venera program, ESA's Venus Express and EnVision missions, PEDRO-V aims to study the performance and durability of materials in the Venusian atmosphere. The goals included gaining insights into fundamental properties and advancing relevant technologies. The Venus mission aimed to explore the planet's atmosphere and chemical composition using miniaturised satellites, enhancing our understanding of material durability in extreme environments. The mission architecture consists of a mothership (carrier spacecraft) and 20 1U CubeSats. The mothership will deliver the CubeSats to Venus and release them into the atmosphere. The mission is designed to leverage the compact and cost-effective nature of CubeSats while maximising scientific return. The requirements of the mission were stated by ESA to provide the overall technical constraints of the development. The CDF team focused on designing the mothership for the mission, with its results detailed in the following section. Once the mission phases and modes of operation were established, each subsystem began its engineering process simultaneously.

**Subsystem Design**

*Trajectory Analysis*: Trajectory analysis is the first step in the design of a space mission as it provides the necessary inputs for the analysis and sizing of other subsystems, such as the cost of the mission in terms of $\Delta V$ and the timeline of the whole mission, taking into account the duration of each orbital manoeuvre. Based on the mission requirements, certain assumptions were made about the launcher (single, optimised burn of the upper stage), the interplanetary phase (Lambert problem) and the target orbit (Hohmann transfer). Iterations of trajectory analysis allowed the results to be checked and refined, for example by considering aerobraking for VOI (Venus Orbit Injection) at the third iteration. The study was divided into a timeline to define the launch window and the subsequent phases: launch injection, interplanetary transfer, aerobraking, transfer to target orbit, orbit propagation around Venus, station keeping and corrections. The team also evaluated Venus's atmosphere with space weather simulations at the time of arrival of the mother spacecraft and assessed the deployment of the CubeSats taking into account eclipse periods, global coverage and communication with ground stations. At the end also planetary protection and orbital decay were examined. The trajectory team primarily collaborated with the propulsion team for $\Delta V$ cost assessments, the C&DH team for global coverage and ground station communication, and the EPS and TCS teams for eclipse evaluations. Providing these inputs to the other teams required the trajectory team to work diligently and quickly at the beginning of the challenge. An important observation was that the trajectory team was proficient in using all the essential mission analysis software, including Ansys STK [13], GMAT [14], and GODOT [15]. This proficiency was especially valuable, as the team had faced challenges with specific subsystem spreadsheets.

*AOCS*: To design the attitude and orbit control system it is necessary to take into account each phase of the mission, the various disturbances in the orbit, but most importantly the functions the satellite must perform. The design of the AOCS was conducted using the Wertz approach [16], defining operational modes, disturbance torques, drafting requirements and iterating the choice of hardware, starting from the inputs provided by the mission phases, trajectory analysis and by the interaction with Thermal, TMTC and Propulsion subsystems. The sizing process was carried out using the provided spreadsheets and considering the worst-case scenario, selecting hardware that was able to compensate for the maximum torque generated by solar radiation pressure and gravitational and atmospheric perturbations. Moreover, given the main objective of the mission, the CubeSats deployment was considered as a disturbance acting on the satellite, representing the highest torque value, which was used to size the attitude thrusters. Being at the early stages of design, perturbations due to the Venus magnetic field were neglected as they are significantly lower than the others.

*Propulsion*: The development of the Propulsion Subsystem is responsible for the final orbit deployment of the spacecraft. It takes as input the $\Delta V$ budget and AOCS requirements, and it gives as an output the propulsion system wet and dry masses as well as the volumes needed. These will affect especially the configurations and structure of the spacecraft. The subsystem design was performed by using the ESA calculation sheets, and references [16, 17]. The $\Delta V$ budget and AOCS requirements influences various aspects such as the nature of propulsion used, tank size, configuration subsystem, overall spacecraft mass and power requirements. This dependency is evident in the mass evolution presented in the results section (see Fig. 2.). Once the mission phases were defined by the Trajectory Analysis team, an initial trade-off was conducted to choose between chemical or electric engines. During the trade-off analysis the team selected a pressure-regulated system to ensure constant thrust for the entire burn time duration and the choice of MON-3/MMH propellants, also used in other Venus missions. This choice was driven by the mission's $\Delta V$ requirement for the VOI, which necessitated high thrust for a relatively short duration and keeping under consideration storability, high-thrust compliance, hypergolic properties, low freezing point and high performances properties of these propellants. Additionally, minor thrust phases were anticipated for correction manoeuvres and drag compensation, which influenced the choice of the number, model, and arrangement of the thrusters. Although the aerobraking phase was not planned until the third iteration, additional bi-liquid thrusters were selected to lower the orbit for insertion during the early iteration phase, requiring heavier tanks and more space in the spacecraft. The final iteration led to the choice of a configuration with four spherical tanks in a $2 \times 2$ arrangement and a central Helium pressurisation tank to minimise CoM shift due to propellants consumption and consequently eliminate potential disturbances to the AOCS subsystem. This configuration also better utilised the space in the spacecraft, which was essential for the Configuration Subsystem. The thermal aspect was also considered by providing the TCS team with the temperature ranges for the tanks.

*TCS*: Approaching the Sun means dealing with rapidly increasing temperatures during daylight. Conversely, when the Sun is not visible, there is a significant drop in temperature, impacting the most sensitive components. The incoming heat sources vary depending on the trajectory of the mothership. In the first iteration, due to the lack of precise trajectory information, the design was based on analyses of other similar missions. Given the proximity to the Sun, the "hot case" was defined as the spacecraft orbiting in sunlight around Venus. This scenario necessitated the design of large radiators, which then led to a temperature drop when the spacecraft orbited the Earth. Typically, the most temperature sensitive components are the electronics; thus, their survivability and operational ranges were derived from existing literature [16]. Using this information and the initial inputs given by the trajectory and configuration subsystems, preliminary calculations for the radiator and heater were made. The first iteration, yield to very large radiators, impacting the configuration team on one side and the heater power needed in the cold phase, deeply affecting the sizes of the battery. Therefore, to lessen the burden on the other subsystems, the TCS team attempted to modify the coatings and enhance the heat transfer through passive components. The decrease of the spacecraft size and the selection of the components with larger operational temperature ranges, made possible through the effort of all the other subsystems, at the end allowed to reduce the power required by the heater and the area of the radiator.

*TMTC & OBDH*: To design the Telemetry and Telecommand (TMTC) subsystem, the team focused on maintaining high performance while minimising mass and power consumption. The first approach involved estimating the link budgets in uplink and downlink with Earth through calculation and datasheet [16]. To this end, the New Norcia's Ground Station was identified, which hosts a 35-metre deep-space antenna that is leveraged for its capabilities to support transmission and reception in both S- and X-band. New Norcia is particularly advantageous due to its strategic location, which offers excellent satellite visibility and radiofrequency clearance ensuring data transmission and reception. Maintaining continuous communication with the Trajectory Analysis team was crucial since the link budget estimates varied with each update they provided regarding the distance between the mothership and the CubeSats. Two antennas were chosen: a Gregorian parabolic antenna as the High Gain Antenna (HGA) and Cassegrain dish antenna as the Medium Gain Antenna (MGA) for safe-mode. Mean contact with Earth is around 4000 seconds per contact so it needs 6 contacts during the mission. For the interlink communication between the mothership and the CubeSats, S-band was selected due to the noisy environment, necessitating robust communication. To achieve this, two half-wavelength dipole antennas as Low Gain Antennas (LGA) were designed to provide an almost isotropic radiation pattern, ensuring Line of Sight (LOS) with the CubeSats. Additionally, these antennas were positioned to facilitate constructive interference. Data from CubeSats are captured at 2 Mb/s with lossless compression, ensuring high-quality data transmission; therefore, QPSK modulation with a half code rate was chosen. On-board data handling (OBDH) is the subsystem which carries and stores data between the various electronics units and the ground segment, via the TMTC subsystem. The system uses cold redundancy for data storage and the main On-Board Computers (OBCs), featuring 32GB for the main OBC and 256GB for data storage. A multi-bus architecture allows easy switching between primary and backup hardware. Command and beacon data use the Controller Area Network (CAN), while AOCS, housekeeping, and camera images use RS485. Payload data acquisition employs multi-drop RS422, and payload and housekeeping data downlink also use RS485. Temperature and deployment sensors use the Inter-Integrated Circuit (I2C) bus.

*EPS*: The development of the power subsystem is relatively straightforward to define for interplanetary missions without requiring extensive technological trade-offs. The high-level design comprised solar arrays, a power distribution and control unit, and a primary battery. During the design process, numerous iterations were necessary, primarily due to the high demands of the thermal control unit and the eclipse durations identified by the mission analysis, which were critical for component sizing. The initial sizing faced significant challenges due to the harsh thermal environment of Venus, leading to high energy requirements. However, through successive iterations, the energy demand decreased substantially. Ultimately, the final design demonstrated strong parallelism with existing Venus missions. Notably, the size of the solar panels was significantly reduced, resulting in a final design of two deployable solar panels, each measuring 2m × 1.4m. This reduction in panel size positively impacted the mechanism and structure of subsystems.

*Mechanisms*: The main objectives regarding the mechanisms of the mission design were the timed, individual ejection of the CubeSats in order to cover the 6 months long operation [18], and the deployment and actuation system of the solar arrays. Later on an additional feature, a camera was implemented to supervise the CubeSat ejections to confirm each satellite stationing with its required safety measures such as a protective box to shelter it from the space debris. The development of the CubeSat ejection system initially aimed to follow a spring actuation with guide rails to minimise space with a wide range of solutions regarding the launch mechanism taking into consideration the environmental and physical differences in Venus and Earth's characteristics, while acknowledging the state of art technology on satellite mechanisms, such as the use of shape memory or Rose's alloys [19]. As this development would have demanded further research and development that were beyond the span of this challenge, the final design was elected to be individual ejection cubes for their established safety and usage in the industry. For the solar array actuation, the motor was sized using the ESA Excel sheet, with a confirmation from the Power team for each iteration. The progress for the concept design was based on the data provided by the team collectively in COMET, as well as using the ECSS recommended standards where applicable. In the final evaluation, a Small Satellite Solar Array Drive Assembly was implemented which constrained the rest of the solar array actuation design. The camera was evaluated through cooperative iterations among the mechanisms, the thermal, the power and the communications teams, with its cover box added later on. SolidWorks [20] was used to model the mechanism concepts in a simplified manner to add later onto the overall CAD.

*Structure and Configuration*: The structure of the mothership was one of the most affected subsystems during the various iterations in terms of changes. Due to the mission requirements, the modification of a subsystem affects the required structure of the spacecraft. The first iteration focused primarily on understanding the strategies used in past missions and some trade-off analysis on material and mass allocation for the structure. This was done based on the expected exploration environment and the linear relationship between payload mass and dry mass in such systems. The work began with a rough estimate of the geometry and masses using the Excel spreadsheet. However, once the configuration started to provide input on the distribution of loads within the mothership, the spreadsheet was abandoned in favour of using a fine CAD (Computer Aided Design) tool to calculate the geometry and masses. The continuous interaction with other teams has resulted in a more compact design, avoiding dead spaces and optimising the inertia matrix as much as possible, as the ratio between empty and filled volume strongly influences the overall mass. In particular, the team interacted with the Configuration, the Thermal and the Propulsion teams to perform a first attempt at FEM (finite element model) analysis. The primary challenge was the absence of a tool for this analysis, necessitating significant effort in modeling the structure to obtain credible results. In the final iteration, the team successfully achieved a mass consistent with the average of past missions with similar payloads, closely aligning with the values from the initial concurrent design sessions. The final structure was made of prepreg + aluminium and reinforced with crossed aluminium beams, for a total mass of about 260 kg. The strategic design in distributing the reinforcement beams enabled the team to securely fix the internal materials with stable support, resulting in Von Mises stress levels significantly below the material limits, approximately 35%. In a similar manner to the structure of the spacecraft, the configuration and placement of each subsystem were highly sensitive to the changes made by different teams in each iteration. To start the design process, certain assumptions needed to be made related to thermal assessments and practical components integration based on the mission's goals. Due to big variations in the required propellant, the propulsion tanks were the largest design driver, initially leading to an oversized spacecraft but ultimately refined to a practical 2m × 2m × 1.2m configuration with a 2 × 2 tank layout to optimize the center of mass and moment of inertia. The second largest design driver for the final configuration of the spacecraft was the thermal requirements as passed on from the TCS team. The total area of the radiators constrained our design largely, and it required many iterations until a value that satisfied the needs of the mission was reached. This had to be done while leaving enough area for the rest of the equipment on the outside of the spacecraft. Finally, the optimal ejection strategy of the CubeSats was calculated to be on the sides perpendicular to flight direction of the spacecraft. Yet due to the existence of the CubeSat mechanism, this face could not be fitted with a large enough radiator. To consolidate this, it was decided that this face would constantly face Venus, only rotating when a CubeSat needed to be ejected – a maneuver comfortably handled by the AOCS system. To design our spacecraft, a tool different from COMET was needed, as it lacks a visualisation aspect. To that end we used initially a properly scaled PowerPoint slide, and finally a CAD model. The final design configuration for the PEDRO-V mothership can be viewed in Fig. 1.

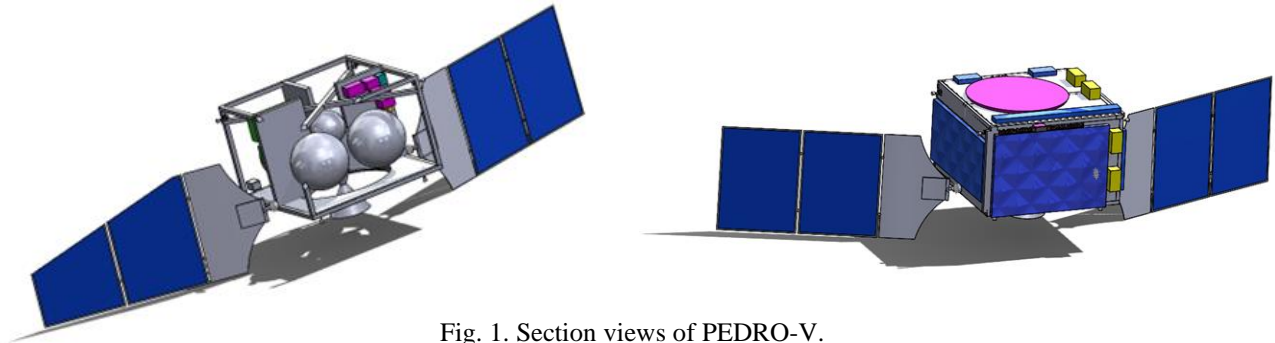

Fig. 1. Section views of PEDRO-V.

**RESULTS AND CONCLUSIONS**

The following picture (Fig. 2.) showcases the mass evolution of spacecraft components over four design iterations, highlighting significant changes in payload, platform, and various subsystems. Initially, from Iteration 0 to Iteration 1, there was a substantial increase in the nominal dry mass at launch, total dry mass, wet mass, and launch mass, reflecting the initial addition of components and adjustments. However, from Iteration 1 to Iteration 2, a notable reduction occurred across all mass categories, indicating optimization and weight-saving measures. This trend of decreasing mass continued into Iteration 3, with further reductions in nominal and total dry mass, while wet and launch masses slightly increased, suggesting a stabilization phase. These fluctuations illustrate the iterative process of refining spacecraft design to achieve an optimal balance between mass and functionality for mission success. During the first iteration, the subsystem teams independently estimated the mass. Although the second iteration saw a temporary increase in mass due to concurrent discussions and integration of inputs from other subsystems, the mass eventually converged to the initial estimates. This alignment reflected values consistent with literature and expert knowledge.

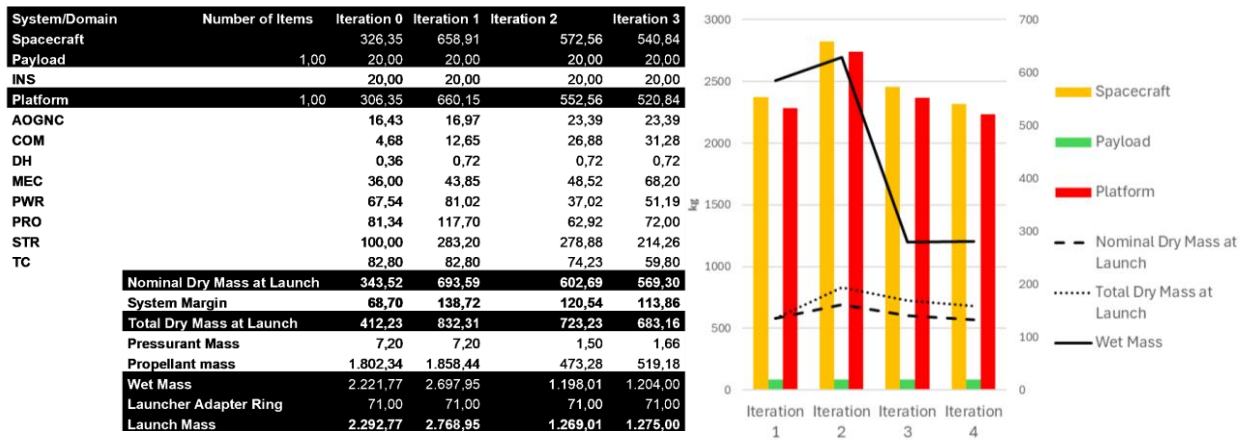

| System/Domain | Number of Items | Iteration 0 | Iteration 1 | Iteration 2 | Iteration 3 |
|---|---|---|---|---|---|
| Spacecraft | | 326,35 | 658,91 | 572,56 | 540,84 |
| Payload | 1,00 | 20,00 | 20,00 | 20,00 | 20,00 |
| INS | | 20,00 | 20,00 | 20,00 | 20,00 |
| Platform | 1,00 | 306,35 | 660,15 | 552,56 | 520,84 |
| AOGNC | | 16,43 | 16,97 | 23,39 | 23,39 |
| COM | | 4,68 | 12,65 | 26,88 | 31,28 |
| DH | | 0,36 | 0,72 | 0,72 | 0,72 |
| MEC | | 36,00 | 43,85 | 48,52 | 68,20 |
| PWR | | 67,54 | 81,02 | 37,02 | 51,19 |
| PRO | | 81,34 | 117,70 | 62,92 | 72,00 |
| STR | | 100,00 | 283,20 | 278,88 | 214,26 |
| TC | | 82,80 | 82,80 | 74,23 | 59,80 |
| Nominal Dry Mass at Launch | | 343,52 | 693,59 | 602,69 | 569,30 |
| System Margin | | 68,70 | 138,72 | 120,54 | 113,86 |
| Total Dry Mass at Launch | | 412,23 | 832,31 | 723,23 | 683,16 |
| Pressurant Mass | | 7,20 | 7,20 | 1,50 | 1,66 |
| Propellant mass | | 1.802,34 | 1.858,44 | 473,28 | 519,18 |
| Wet Mass | | 2.221,77 | 2.697,95 | 1.198,01 | 1.204,00 |
| Launcher Adapter Ring | | 71,00 | 71,00 | 71,00 | 71,00 |
| Launch Mass | | 2.292,77 | 2.768,95 | 1.269,01 | 1.275,00 |

Fig. 2. Overall view on mass evolution (credits: F. Rometsch, R. Biesbroek and D. Wischert, ESA/TEC-SYE)

**Educational Impact**

Integrating CE into educational programs, such as the ESA Academy's Concurrent Engineering Challenge, prepares future aerospace engineers for complex space exploration tasks [21]. CE represents a significant evolution from traditional sequential methods, offering a more efficient and collaborative approach to space mission design. This approach was particularly beneficial during the challenge, where students from diverse backgrounds quickly adapted to a fast iteration approach, learning new skills through hands-on experience. Some of the students were used to agile development, focusing on short cycles of iterative progress, while others preferred approaches that prioritize precise outcomes with lower uncertainties. During the development of the mission, the absence of a streamlined approach and the diversity of students' methodologies initially posed a challenge. The team rapidly acquired new skills and knowledge through this experience, exhibiting remarkable flexibility and adaptability in the face of daily deadlines. However, with limited time, the team adapted quickly, learning by doing. By the end of the week, a fast iteration approach emerged as the consensus, demonstrating the core value of CE: fostering rapid development cycles. This convergence within a week was driven by the time constraints and the collaborative efforts of team members. Having a ready-to-use software environment was crucial for the concurrent study, though integrating such tools into typical student settings is often challenging due to complexity and learning curves. COMET proved invaluable for the challenge, enabling efficient collaboration and mission design. However, using it in later phases may pose challenges due to the team's limited familiarity with its features. Additionally, COMET has inherent limitations, such as the inability to directly design systems or perform calculations without integration with tools like Excel. The CDF at ESEC-Galaxia along with the comprehensive support from CDF engineers and ESA education staff, was crucial in allowing the team to focus on the engineering aspects of the challenge. This supportive environment facilitated effective learning and the practical application of CE principles, ultimately leading to the successful completion of the PEDRO-V mission proposal.

The ESA Academy's Concurrent Engineering Challenge 2024 provided a unique educational experience, blending theoretical knowledge with practical application. By working on a real-world project under professional guidance, students gained a deep understanding of the complexities and interdisciplinary nature of space mission design. The challenge fostered critical skills such as teamwork, problem-solving, and effective communication, essential for future aerospace careers. The event also underscored the importance of active learning methodologies, such as PBL, in enhancing student engagement and motivation. Students directly applied their academic knowledge to a tangible project, resulting in a more enriching and impactful learning experience. Despite the challenges, all groups met the mission requirements and constraints, with each design featuring strong and unique elements. Participants agreed that concurrent engineering is a highly effective approach for both professional and academic settings. The challenge provided valuable insights from ESA experts and colleagues from other universities, all working towards a common goal.

**Future Directions**

Given the growing importance of CE in business success, it is essential that engineering students understand and know how to effectively apply this concept. The successful implementation of the PEDRO-V mission concept demonstrates the potential of concurrent engineering in fostering innovative solutions and validating complex mission concepts in a limited amount of time. Future research could focus on refining the mission of PEDRO-V in order to reach a higher design level with a CE approach and exploring the long-term impact of active learning approaches on student engagement and STEM education outcomes. Additionally, CE could be tested in more advanced mission phases beyond feasibility studies.

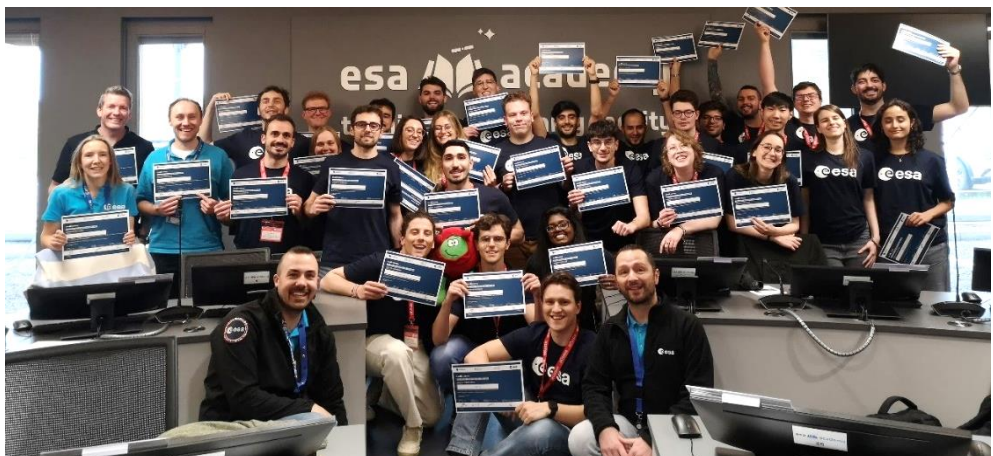

Fig. 3. The CEC 2024 team (credits: ESA Academy)

**Acknowledgements**

The authors sincerely thank ESA System Engineers Robin Biesbroeck, Daniel Wischert, and Flavie Rometsch for their invaluable assistance, and extend their appreciation to Eleftherios Karagiannis, Jaime Carà Junior, and Ariane Dedeban for organizing the event and ensuring a memorable challenge week. For space matter, the affiliation of the authors are here reports, also to highlight the multicultural fostering of the initiative. A special thanks to: the Univeristy of Naples "Federico II", the Instituto Superior Técnico of Lisbon, the Politecnico of Turin, the Politecnico of Milan, the National University Of Science And Technology Politehnica of Bucharest, the Newcastle University, the Technical University of Munich, the University of Roma Tre, the Rzeszow University of Technology, the Technical University of Denmark (DTU), and the Technische Universität of Berlin.